\documentclass[prb,preprint]{revtex4-1} 

\usepackage{amsmath}  
\usepackage{amsfonts} 
\usepackage{graphicx} 

\newcommand{\Eins}
\newcommand{\fref}[1]{Fig.~\ref{#1}}

\newcommand{\eref}[1]{Eq.~(\ref{#1})}
\newcommand{\esref}[2]{Eqs.~(\ref{#1}) and (\ref{#2})}

\newcommand{\cref}[1]{chapter~\ref{#1}}

\newcommand{\Cref}[1]{Chapter~\ref{#1}}

\usepackage{ulem}  
\usepackage{color} 
\definecolor{gray}{rgb}{0.5,0.5,0.5}

\begin{document}

\title{Wave packet propagation and the materialization of classical trajectories}    

\author{James M.\ Feagin}
\affiliation{Department of Physics, California State University-Fullerton, Fullerton, CA 92834, USA}
\author{John S. Briggs}
\affiliation{Institute of Physics, University of Freiburg, Germany \\
and Royal University of Phnom Penh, Cambodia}

\begin{abstract}

Unbound wave packets propagating to macroscopic  space and time coordinates become proportional to their (Fourier transform) momentum distribution at earlier times whereby the asymptotic coordinates and the initial momenta are connected everywhere by appropriate classical trajectories. This asymptotic \emph{ imaging theorem} is relevant to every quantum reaction involving macroscopic extraction and detection of fragments emerging from a microscopic volume. It  justifies the usual assumption of classical particle motion used in the design of particle detectors.  We illustrate this quantum to semiclassical transition with the example of the free propagation of a wave packet in one dimension, a standard problem treated in introductory lectures on quantum mechanics. We indicate generalizations appropriate for more advanced discussions.

\end{abstract}


\maketitle

\section{Introduction}

In very many quantum mechanics text books, the quantum to classical transition is indicated by the requirement that 
wave packets remain spatially localized. This condition that a wave packet can be associated with a single localized particle is due originally to Schr\"odinger in 1926.\cite{Schr} The proof in 1927 by Kennard \cite{Kennard} and Ehrenfest \cite{Ehren} that the  average values of the wave packet position follows Newton's Law (at least for a certain class of potentials) is then used to cement this localization as the condition for quantum particles to manifest classical behavior. 

Hence, the spreading in space of general wave packets, pointed out by Heisenberg \cite{Heisenberg} and Kennard \cite{Kennard} in 1927, is viewed in text books as a signature of \emph{ quantum}, i.e.\ explicitly non-classical, behavior. However, Kennard also pointed out that the spreading in space mirrors exactly the free classical motion of an ensemble of particles of different  momenta. Here, in support of Kennard, we will show explicitly that the spreading is actually a mirror of {\emph{classical} motion and signifies the appearance of classical trajectories within the quantum wave function.
Although, as shown by Ehrenfest,\cite{Ehren} his theorem is satisfied by the spreading wave packet, the averages defined in the theorem have little relevance for wave packets spread over cubic metres of space, as is applicable to particle detectors. }

That the momenta $p$ appearing mathematically from the Fourier transform (FT) of a spatial wave function in $x$ at time $t$ correspond, after propagation to large distance on a macroscopic scale, to the classical momenta defined by $p = mx/t  \equiv mv$, is due to Kemble in 1937.\cite{Kemble} This somewhat neglected section of his text, particularly considering the light it sheds on the quantum to classical transition, forms the basis of the developments presented here. Kemble showed that, within the preservation of the quantum wave function, asymptotically the spatial and momentum  \emph{coordinates} are connected by classical trajectories. He showed that the spatial wave function becomes proportional to the momentum wave function at earlier times. This result has become known subsequently as the \emph{imaging theorem} (IT).

In most modern experiments, particles (molecules, atoms, nuclei) are tracked from a microscopic reaction volume onto detectors placed at macroscopic distances.
The assumption that the detected particle can be taken as moving classically has been largely absorbed into the everyday practice of experimenters.
Even for light particles such as electrons, their path to the detector is assigned to a classical trajectory and the time of flight is used to infer values of dynamical properties.
Despite this assumption of classical motion, coincident detection of quantum correlation between fragments, indicates the preservation of a many-particle wave function out to asymptotic distances. The IT, which shows that classical trajectories materialize autonomously within the quantum wave function,
provides the explanation of this dichotomy and deserves wider recognition in elementary quantum theory.

The concentration on time-of-flight (TOF) measurements in modern experiments can be connected only indirectly to the scattering theory presented in many introductory courses. There, interest is concentrated often on time-independent scattering theory giving scattering amplitudes and cross-sections. 
Such a theory is inappropriate for example, for modern experiments using ultra-short laser pulse fragmentation. We will show how the IT is eminently suited to describe the collection and interpretation of  particle-counting data based on TOF measurements.

Although the propagation in time of coordinate and momentum wave functions is a standard feature of introductory courses, the insight that the asymptotic IT brings to
the interpretation of quantum and classical behavior, usually neglected, deserves a place in such courses. The motion of quantum particles and their detection macroscopically, embodied in the IT result, is fundamental to all types of basic collision experiments. 

In the next section, we introduce a time-dependent one-dimensional (1D) gaussian wave packet freely moving and examine its asymptotic form to consider detection at macroscopic distances and times. The result introduces the IT for free propagation and is the main point of the paper. We further compare momentum- and coordinate-space propagation  and, using semiclassical analysis, generalize the IT for free propagation to arbitrary initial wave packets. In section III, we introduce additional details of the semiclassical description and present a statistical simulation of particle detection and initial wave packet reconstruction using the IT. In the appendices we extend the IT to unbound propagation in external fields along with a derivation in stationary phase approximation of the free-propagation IT in momentum space for an arbitrary initial wave packet. 

Analytical and computational details presented in this paper are provided online in a working \emph{Mathematica} notebook. \cite{mathematica}

\section{The freely-moving gaussian wave packet and the imaging theorem}

The asymptotic behavior of a moving wave packet in coordinate space is well illustrated with the standard form of a gaussian wave packet. In one dimension this has the spatial distribution at some arbitrary initial time $t = 0$
\begin{equation}
\label{xwfn0}
\Psi(x, t = 0) = \frac{1}{(\pi\sigma^2)^{1/4}}\, \exp{\left(-\frac{x^2}{2\sigma^2}\right)},
\end{equation}
that is, a gaussian of width $\sigma$ centered on the origin $x=0$.

The wave packet in momentum space at time $t = 0$ is obtained as the (inverse) Fourier transform of \eref{xwfn0} and reads
\begin{equation}
\label{pwfn0}
\begin{split}
\tilde\Psi(p,0) &= (2\pi \hbar)^{-1/2} \int  \Psi(x, 0) \, e^{-i p x/\hbar}  \, dx \\
	&= \left(\frac{\sigma^2}{\hbar^2\pi}\right)^{1/4} e^{- \sigma^2p^2/(2\hbar^2)}
\end{split}
\end{equation}
with width given by $\hbar/\sigma$. 
The corresponding probability distributions $|\Psi(x, 0)|^2$ and $|\tilde \Psi(p, 0)|^2 $ are both normalized to unity.

\subsection{Momentum-space propagation}

In accordance with conservation of momentum, the momentum wave packet propagates unchanged in shape, acquiring simply an energy-time phase factor upon propagation to finite times,
\begin{equation}
\label{Pprop}
\tilde \Psi(p,t) = \tilde \Psi(p,0)\, \exp{\left(-\frac{i}{\hbar}\frac{p^2}{2m}t\right)}.
\end{equation}

Using \eref{pwfn0} it is convenient to express this wave function in the form
\begin{equation}
\tilde\Psi(p,t) = \left(\frac{\sigma^2}{\hbar^2\pi}\right)^{1/4} \exp{\left(- \frac{ \sigma_t^2\,p^2}{2\hbar^2}\right)},
\end{equation}
where we introduce the complex-valued width $\sigma_{t} \equiv \sigma(1 + i\tau)^{1/2} $
with characteristic time scale $T \equiv m\sigma^2/\hbar$ and $\tau \equiv t/T$. 

Propagation of the coordinate space wave packet to finite times is readily evaluated with a Fourier transform which gives
\begin{equation}
\label{xwfnt}
\begin{split}
\Psi(x,t) &= (2\pi \hbar)^{-1/2} \int  \tilde \Psi(p,t) \, e^{i p x /h} \, dp  \\
&= \frac{1}{(\pi\sigma_t^2)^{1/4}}\, \exp{\left(-\frac{x^2}{2\sigma_t^2}\right)},
\end{split}
\end{equation}
also a normalized gaussian centered on the origin $x=0$ but now with a time-dependent complex width $\sigma_t$. 

\begin{figure}[t]
\includegraphics[scale=.2]{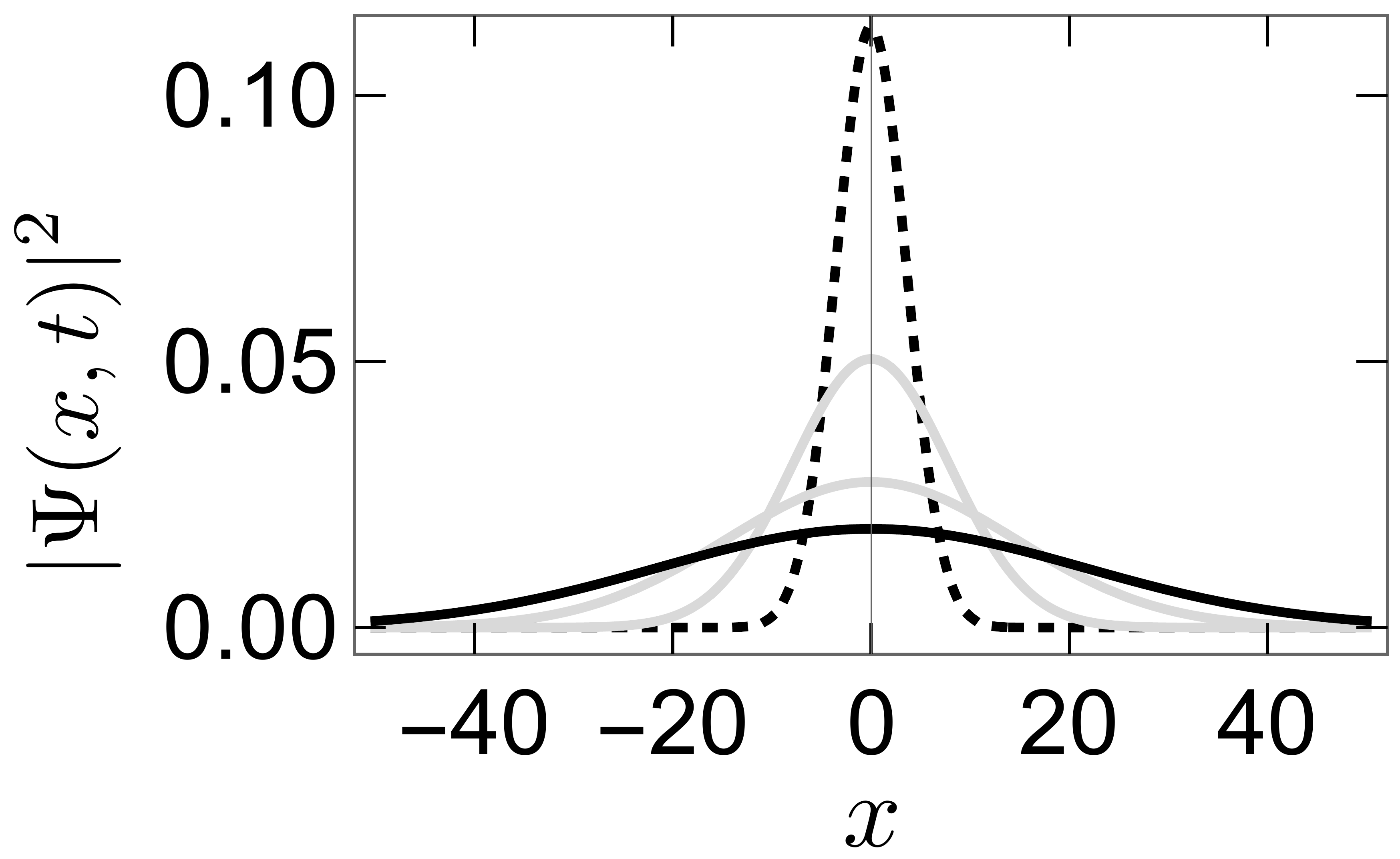}
\caption{\label{fig1a} Free propagation of a 1D gaussian wave packet with time for $\hbar = m = 1, \sigma = 5$ showing $|\Psi(x,t)|^2$ as a function of $x$  for four different times $t = 0, 50, 100, 150$. The dotted curve is the initial density, the black curve the final.   }  
\end{figure}

To preserve normalization as the wave packet expands in time, the normalization factor $(\pi\sigma_t^2)^{-1/4}$ is time dependent. This expansion can be seen explicitly by writing the propagating wave packet in the equivalent form
\begin{equation}
\label{exact1d}
\begin{split}
\Psi(x,t) &= \frac{1}{[\pi\sigma^2(1 + \tau^2)]^{\frac{1}{4}}} \exp{\left[-\frac{i}{2}\arctan{\tau}\right]}\\
&\times \exp{\left[-\frac{x^2}{2\sigma^2(1+\tau^2)}\right]}\, \exp{\left[i\frac{x^2\,\tau}{2\sigma^2(1+\tau^2)}\right]}.
\end{split}
\end{equation}
The width of the gaussian is now $\sigma\,(1 + \tau^2)^{1/2}$, which increases proportional to $\tau$ asymptotically.
This expansion in time of $|\Psi(x,t)|^2$ is illustrated in Fig.\ \ref{fig1a}. 

We emphasize that there is no overall translation of the wave packet $\Psi(x,t)$. It simply expands in time and mimics the release of fragments from a quantum reaction. Some fragments will have a large momentum but a small probability $|\tilde \Psi(p,0)|^2$ for release, while some will have small momenta with relatively larger probabilities. 

To simulate macroscopic detection, we extract the asymptotic form of the expanding wave packet at large distances and times relative to the extent of the initial wave packet, that is we take $x \gg \sigma$. 
When viewed from a detector at macroscopic $x$, the initial wave packet is of negligible extent $\sigma \ll \sigma_{t}$.
Expanding the exponent in \eref{xwfnt} about $\sigma = 0$ ($T \ll t$) gives
\begin{equation}
	-\frac{x^2}{2\sigma_t^2} \approx i \frac{m x^2}{2\hbar t} - \frac{m^2 x^2 \sigma^2}{2\hbar^2 t^2} + \mathcal{O}(\sigma^{3}).
\end{equation}
Similarly expanding the normalization factor to $\mathcal{O}(\sigma)$, one obtains
\begin{equation}
\label{PsiFreeIT}
\begin{split}
\Psi(x, t) &\approx \left(\frac{m}{ i t}\right)^{1/2} \exp\left[i \frac{m x^2}{2\hbar t}\right] \times \left(\frac{\sigma^2}{\pi \hbar^2}\right)^{1/4} \exp\left[ - \frac{(m x/t)^2 \sigma^2}{2 \hbar^2}\right]\\
&= \left(\frac{m}{i t}\right)^{1/2} \exp\left[i \frac{m x^2}{2\hbar t}\right] \tilde \Psi(p,0),
 \end{split}
\end{equation}
where the classical momentum $p = m x/t$ has been introduced.  

\begin{figure}[b]
\includegraphics[scale=.2]{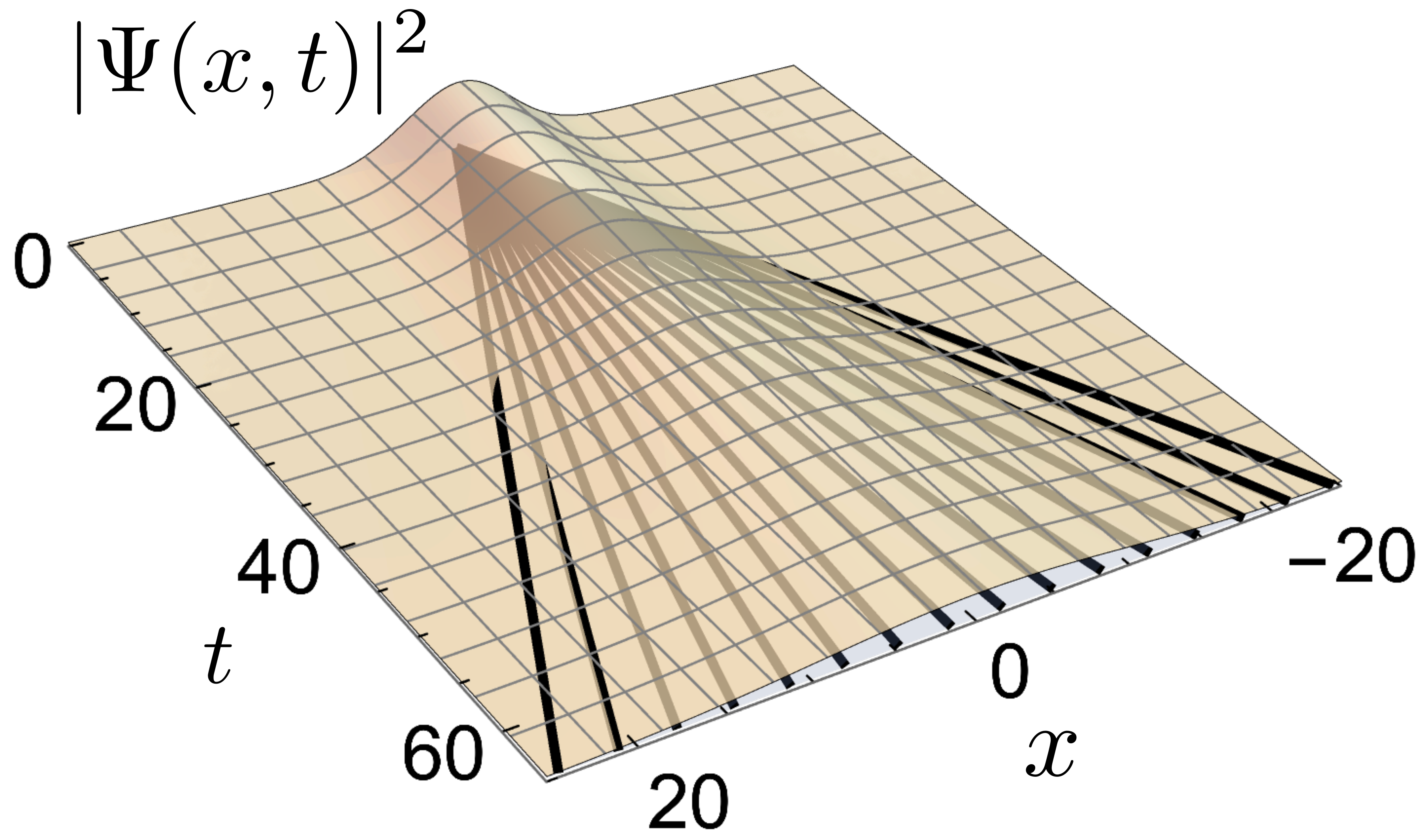}
\caption{\label{fig1b} Free propagation of a 1D gaussian wave packet with time for $\hbar = m = 1, \sigma = 5$ showing $|\Psi(x,t)|^2$ as a function of $x$ and $t$. The straight lines radiating out from the origin are classical free trajectories $x(t) = p t/m$ for a range of $p$.  }  
\end{figure}

\eref{PsiFreeIT} is the 1D \emph{imaging theorem} (IT) for free asymptotic propagation. Important is that the amplitude of the wave function at each final $x$ is connected by a classical trajectory to a unique fixed initial momentum $x = p t/m$. There is no longer an integral over all $p$ values as in \eref{xwfnt}. The transition to classical mechanics occurs in the arguments of the wave functions. For example, the spreading of the wave function in time is linked to the natural separation in time of classical trajectories with different initial momenta $p$ depicted in Fig.\  \ref{fig1b}. The wave functions themselves are preserved giving rise to possible quantum effects such as interference and diffraction. 
This mixed quantum-classical character is the hallmark of the IT. \cite{BrF} 

\eref{PsiFreeIT} is a semiclassical limit with asymptotic normalization $(m/t)^{-1/2}$ defined by the classical probability $dp/dx = m/t$ for a bundle of trajectories in the range $dp$ about $p = m x/t$ to reach $x$ in the range $dx$. In the IT limit, the quantum probability density for detection at $x$ is then
\begin{equation}
\label{PsisqIT}
 |\Psi(x, t)|^2  \approx \frac{dp}{dx} \, |\tilde \Psi(p, 0)|^2.
\end{equation}
This form has a fully classical, statistical interpretation. An ensemble of particles with probability density $|\tilde \Psi(p, 0)|^2$ of initial momentum $p$ move on classical trajectories with the classical density $dp/dx$ and are imaged at later times as the position probability density  $|\Psi(x, t)|^2$. Quantum mechanics furnishes the initial momentum distribution.

\begin{figure}[t]
\includegraphics[scale=.18]{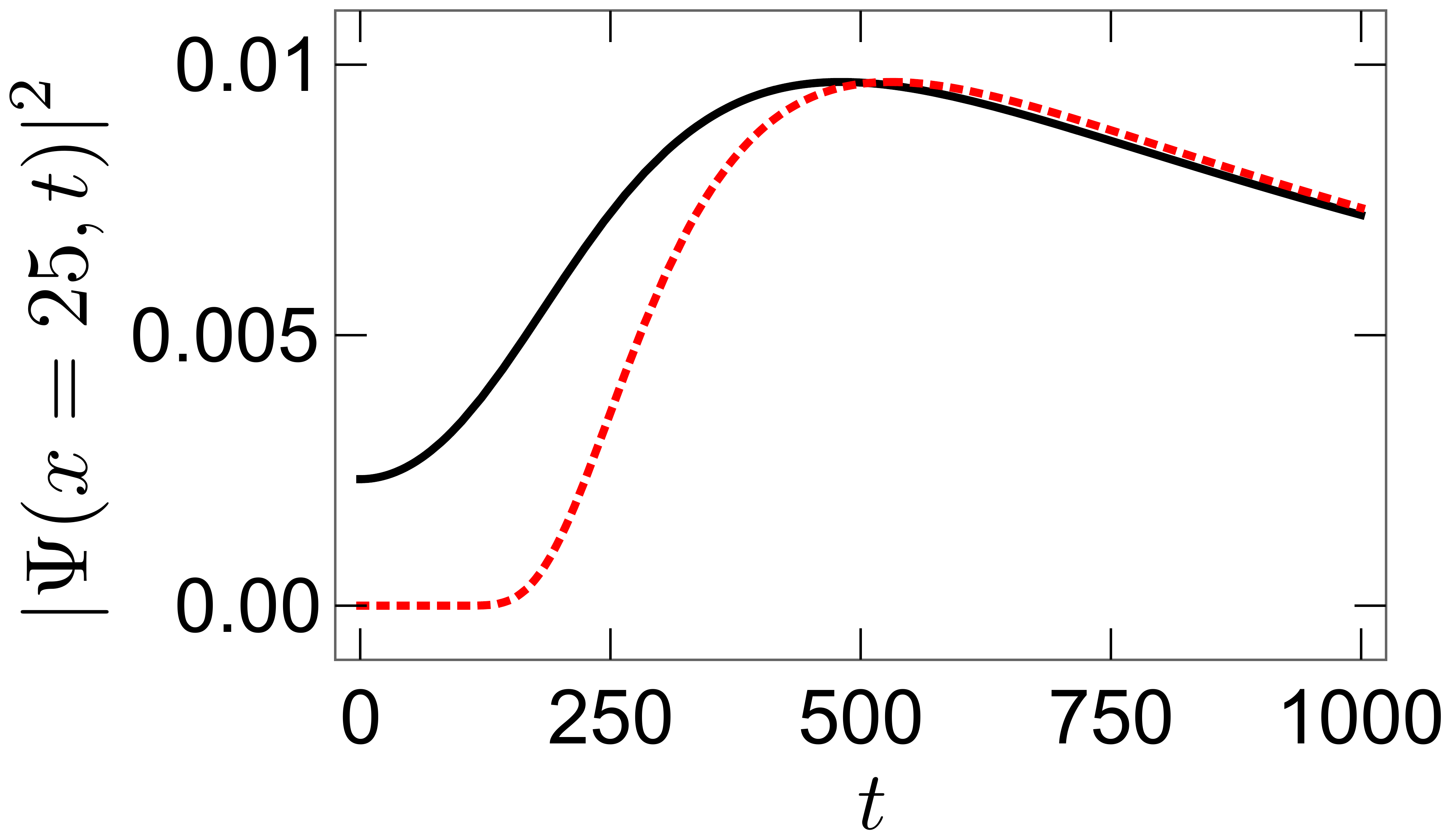}
\caption{\label{fig2} Free propagation of a 1D gaussian wave packet with time in a.u.\ for a \emph{fixed} detector position $x = 25 \, \, \mbox{a.u.}$\ with $\sigma = 10$ so that $T = 100$. The solid (black) curve shows the exact  quantum probability distribution $|\Psi(x,t)|^2$ of detector hits with time. The dotted (red) curve is the IT limit from \eref{PsisqIT}.  }
\end{figure}

In \fref{fig2} we show the probability density as a function of time for a particle to reach a detector placed at $x = 25$ in atomic units (1 a.u.\ $\approx 5.29 \times 10^{-11} \, \mbox{m}$) defined by $\hbar = m_e = 1$, where $m_e$ is the mass of an electron. One sees a rise as the wave packet moves out but then a fall as the wave packet spreads. The IT converges to the exact result very quickly on the atomic scale ($\mbox{1 a.u.}\sim10^{-17} \, \mbox{s}$) and then the probability decreases proportional to $1/t$ in line with \eref{PsiFreeIT}.  

The IT limit \eref{PsiFreeIT} emphasizes propagation out to macroscopic distances compared to the extent of the initial wave packet, i.e.\ $x \gg \sigma$.\cite{validity} 
We highlight this emphasis further in the next section.

\subsection{Coordinate-space propagation}

It is instructive to re-derive the IT by propagating directly in coordinate-space rather than in momentum space with \eref{xwfnt}.
One introduces the quantum coordinate propagator $K(x, x';t)$ to connect $ \Psi(x,0)$ with $ \Psi(x,t)$ according to
\begin{equation}
 \label{Kpropagation}
\Psi(x,t) =  \int K(x, x';t) \, \Psi(x',0) \, dx',
\end{equation}
where in the case of free motion, \cite{GottfriedNEW}
\begin{equation}
 \label{Kfree}
K(x, x';t) = \left(\frac{m}{2\pi i \hbar t} \right)^{1/2} \exp \left[i \frac{m}{2\hbar t} (x-x')^2 \right].
\end{equation}
The integral in \eref{Kpropagation} is readily evaluated with \eref{Kfree} and the initial gaussian wave packet \eref{xwfn0}. The exact quantum result \eref{xwfnt} is obtained as before. 
One also sees that the IT limit \eref{PsiFreeIT} can be also expressed in terms of the propagator as
\begin{equation}
\label{PsiFreeITK}
\Psi(x, t) \approx (2\pi \hbar)^{1/2} K(x, 0;t) \, \tilde \Psi(p,0).
\end{equation}

The free-motion propagator \eref{Kfree} facilitates a simple and direct generalization of the IT to an \emph{arbitrary} initial wave packet $\Psi(x',0)$ in \eref{Kpropagation}. Assuming only that $\Psi(x',0)$ is confined to a small range of $x'$ compared to the asymptotic spread of $\Psi(x, t \rightarrow \infty) $, one can expand the exponent in $K(x, x';t)$ about $x'=0$ and obtain to lowest order in $x' \ll x$ (with $\exp[i m x'^2/2\hbar t] \sim 1$)
\begin{equation}
 \label{KfreeApprox}
\begin{split}
K(x, x';t) &\approx K(x, 0;t) \exp \left[\frac{i}{\hbar} \left(m \frac{x}{t} \right) x' \right] \\
		&\approx K(x, 0;t) \exp \left[-i p x'/\hbar \right],
\end{split}
\end{equation}
which again introduces the classical initial momentum $p = m x/t$.
With this approximate expression, the integral in \eref{Kpropagation} evaluates directly as an inverse Fourier transform and gives \eref{PsiFreeITK} but now generalized to an arbitrary initial wave packet.
We generalize this result further to unbound motion in external fields in the Supplementary Material. \cite{ITF} 

The momentum integral in \eref{xwfnt} can also be evaluated approximately for an \emph{arbitrary} initial momentum wave packet $\tilde \Psi(p,0)$ to derive the asymptotic IT limit.  As $t \rightarrow \infty$, the exponent in \eref{xwfnt} defines a rapidly oscillating phase $\phi = p x - p^2 t/2m$. Asymptotically then, the integral has nonvanishing values only for phases near the stationary value defined by $d\phi/dp = x - p t/m \equiv 0$. That is, the integral is nonvanishing only near $p$ values defined by the classical trajectories, $p \rightarrow m x/t$. Evaluating the integral in stationary-phase approximation gives directly the IT limit \eref{PsiFreeIT} for free propagation of an arbitrary initial wave packet. 
We present details in the Supplementary Material. \cite{ITSPA}

\section{Additional semiclassical aspects}

We emphasize the semiclassical nature of the propagation by noting that the quantum propagator \eref{Kfree} is also the exact semiclassical propagator,  \cite{GottfriedNEW, Gutz}
\begin{equation}
 \label{Ksc}
K_{sc}(x, x';t) = \left(\frac{1}{2\pi i \hbar} \right)^{1/2} \left|-\frac{\partial^2S_c}{\partial x\partial x'} \right|^{1/2} e^{i S_c(x,x';t)/\hbar},
\end{equation}
where $S_c(x,x';t) = m (x-x')^2/2t$ is the classical action for free motion such that the classical momenta are defined by
\begin{equation}
 \label{classicalmom}
p' = -\frac{\partial S_c}{\partial x'} = \frac{m (x-x')}{t} = \frac{\partial S_c}{\partial x} = p,
\end{equation}
as expected for free motion. 
The normalization mixed derivative
\begin{equation}
\label{VV1D}
-\frac{\partial^2S_c}{\partial x\partial x'} = \frac{\partial p'}{\partial x} = \frac{m}{t} = \frac{dp'}{dx}
\end{equation}
is the 1D Jacobian needed for the change of variables $x$ to $p'$ as in  \eref{PsisqIT}.
In higher dimension, \eref{VV1D} is known as the Van Vleck determinant and was introduced by Van Vleck as early as 1928 in his consideration of the connection of quantum mechanics to classical mechanics. \cite{VV}

\subsection{Asymptotic eigenfunction of the classical momentum}

Application of the momentum operator $\hat p$ to the free propagator \eref{Kfree} gives
\begin{equation}
\hat p \,K = -i\hbar \frac{\partial K}{\partial x} = \frac{m(x-x')}{t} \,K
\end{equation}
One sees immediately from \eref{PsiFreeITK} that the asymptotic wave function in the IT limit is locally an eigenfunction of  momentum with eigenvalue the classical momentum. That is, as $t \rightarrow \infty$,
\begin{equation}
\label{efnp}
\hat p \,\Psi(x,t) \approx  \frac{mx}{t} \Psi(x,t) = p\,\Psi(x,t).
\end{equation}
It follows that with macroscopic TOF measurements $x$ and $p = m x/t$ can be determined simultaneously to arbitrary accuracy in strong violation of the Heisenberg Uncertainty Principle.  That is, given a spatial uncertainty $\delta x$, introduced for example by the finite width of a detector channel, the corresponding momentum uncertainty $\delta p \approx m \delta x/t = p \,  \delta x/x$ can be made arbitrarily small with macroscopic $t$ and $x$ values, so that $\delta x \, \delta p \approx m \delta x^2/t = p \, \delta x^2/x \ll \hbar$.
This is just another basic characteristic of macroscopic wave packet propagation.

\subsection{Particle counting and wave packet reconstruction}

The IT ensures that all information regarding a quantum reaction process at an earlier time is encoded in the arrival-time distribution, or time spectrum, of reaction fragments arriving at a remote detector located at fixed $x$. This distribution is proportional to $|\Psi(x,t)|^2$ as a function of the TOF $t$ for $x$ fixed, as depicted in Fig.\ \ref{fig2}, and the IT allows us to reconstruct the initial momentum distribution $|\tilde \Psi(p,0)|^2$.

For free motion, substituting $t \rightarrow m x/p$ from the classical trajectory and dividing by the classical density $dp/dx = m/t$, we obtain from \eref{PsisqIT} for fixed $x$ 
\begin{equation}
\label{phiITsq}
|\tilde \Psi(p,0)|^2 \approx  \frac{t}{m} \, |\Psi(x,t)|^2   \, 
	\rule[-2mm]{.1mm}{6mm}_{\,t \rightarrow m x/p}.
\end{equation}
This IT approximation becomes exact for $x$ and $t$ large enough and certainly for the macroscopic parameters of a typical laboratory apparatus.     

In this sense, the spatial wave function reaching a macroscopic detector images the momentum wave function emanating from a microscopic reaction at earlier times. Of course, the images are derived from the statistical ensemble of particle detector hits at random arrival times $t$. 

To illustrate, consider a 1D quantum harmonic oscillator in its ground state (also a gaussian) with natural frequency $\omega$ and energy $\hbar \omega/2$. At $t=0$ some external mechanism turns off the oscillator's restoring force, releasing the mass $m$ to depart the origin unbounded. Its initial coordinate and momentum wave functions are given by \esref{xwfn0}{pwfn0} with the replacement $\sigma^2 \rightarrow \hbar/m \omega$. 
 
 Let the particle be an electron with an initial harmonic-oscillator  energy of $1.36 \, \mbox{eV}$ so that $\omega = 0.1 \, \mbox{a.u.}$ (1 a.u.\ of frequency $\approx 4.13 \times 10^{16} \, \mbox{s}^{-1}$). 
 Electron detachment to unbound motion could be for example photoionization by a fast laser pulse at $t=0$.
 To simulate a real experiment, we consider a detector at the macroscopic distance $x= +30 \, \mbox{cm}$.  \cite{FBr2} 
 The width of the momentum distribution  $\hbar/\sigma = \sqrt{m \hbar \omega}$ gives an estimate of the average initial momentum, $\bar{p} = \sqrt{m \hbar \omega} = 0.32 \, \mbox{a.u.}$, so that the average TOF classically is $\bar{t} = m x/\bar{p} =  0.43 \, \mu\mbox{s}$. 
Moreover, the characteristic time scale that determines the transition to asymptotic times is now simply proportional to the period of the oscillator, $T = \omega^{-1} \approx 2.4 \times 10^{-10} \, \mu\mbox{s}$, so that the IT limit $t \gg T$ is well satisfied.

 \begin{figure}[b]
\includegraphics[scale=.25]{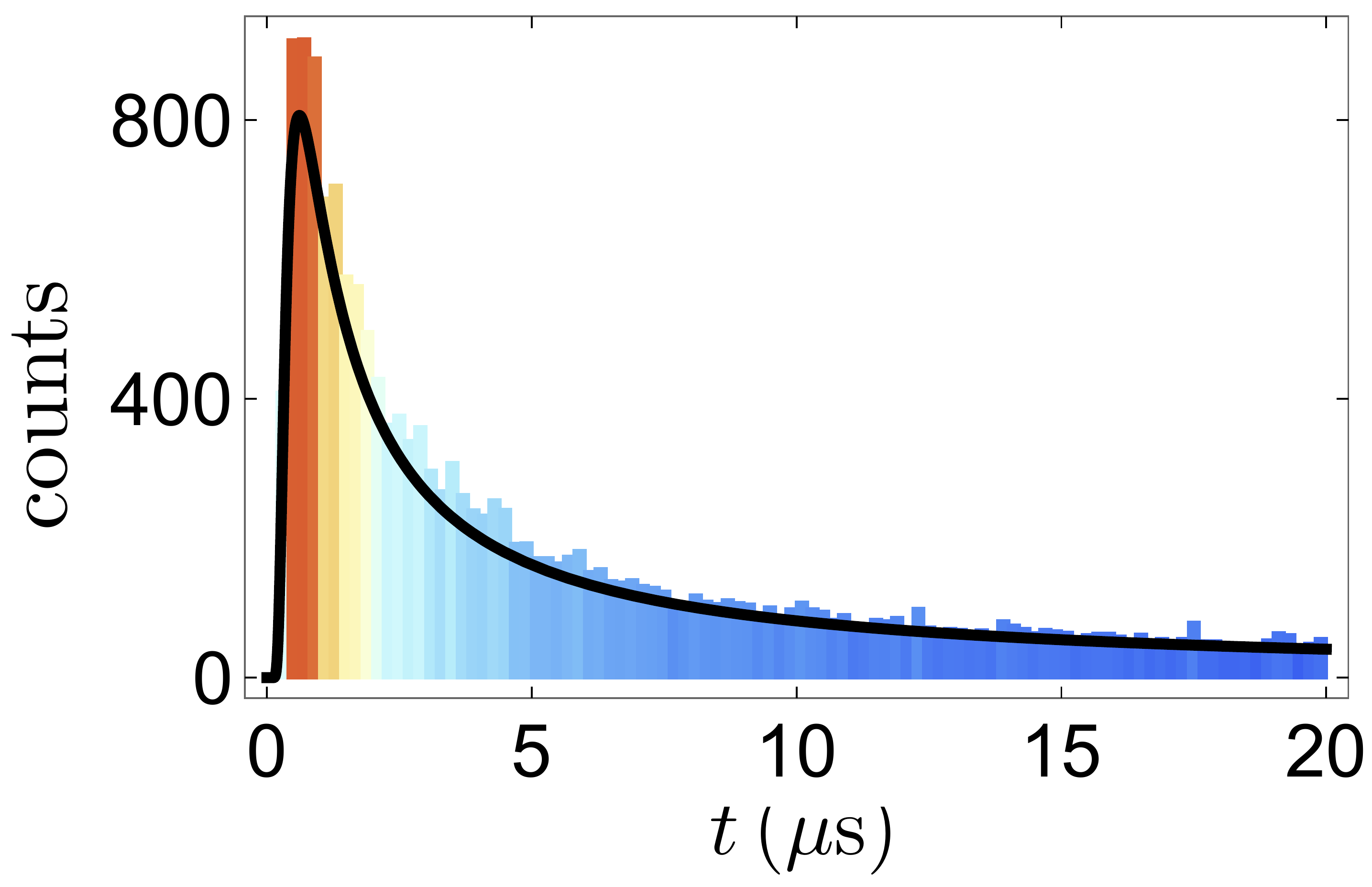}
\caption{\label{fig3a} Histogram of $10^5$ simulated electron detections for fixed $x = 30 \, \mbox{cm}$ with bin widths of $0.2 \, \mu\mbox{s}$. The simulation is based on the freely propagated $n=0$ harmonic oscillator distribution $|\Psi(x,t)|^2$ from \eref{xwfnt} with $\sigma^2 \rightarrow \hbar/m \omega$ and $\omega = 0.1 \, \mbox{a.u.}$ The solid curve shows $|\Psi(x = 30 \, \mbox{cm},t)|^2$ in units of $\mbox{cm}^{-1}$ scaled by $10^5$. }
\end{figure}

 We use rejection sampling \cite{rejectsample} to simulate $10^5$ repeated electron detachment and TOF detection events. The resulting list of random arrival-time values $t$ are distributed according to $|\Psi(x,t)|^2$ from \eref{xwfnt} with $\sigma^2 \rightarrow \hbar/m \omega$ in units of $\mbox{cm}^{-1}$. The list simulates actual random detector hits over a macroscopic time interval. 
The full details of the computation are presented in a working \emph{Mathematica} notebook provided online. \cite{mathematica}

Fig.\ \ref{fig3a} shows a histogram of the simulated random arrival-time detections of electrons (time spectrum for fixed $x$). The time axis is given in microseconds ($\mu$s) with bin widths of  $0.2 \, \mu\mbox{s}$, readily achievable experimentally.
In this statistical sense we see that the wave function fully survives propagation to macroscopic distances.

\begin{figure}[t]
\includegraphics[scale=.25]{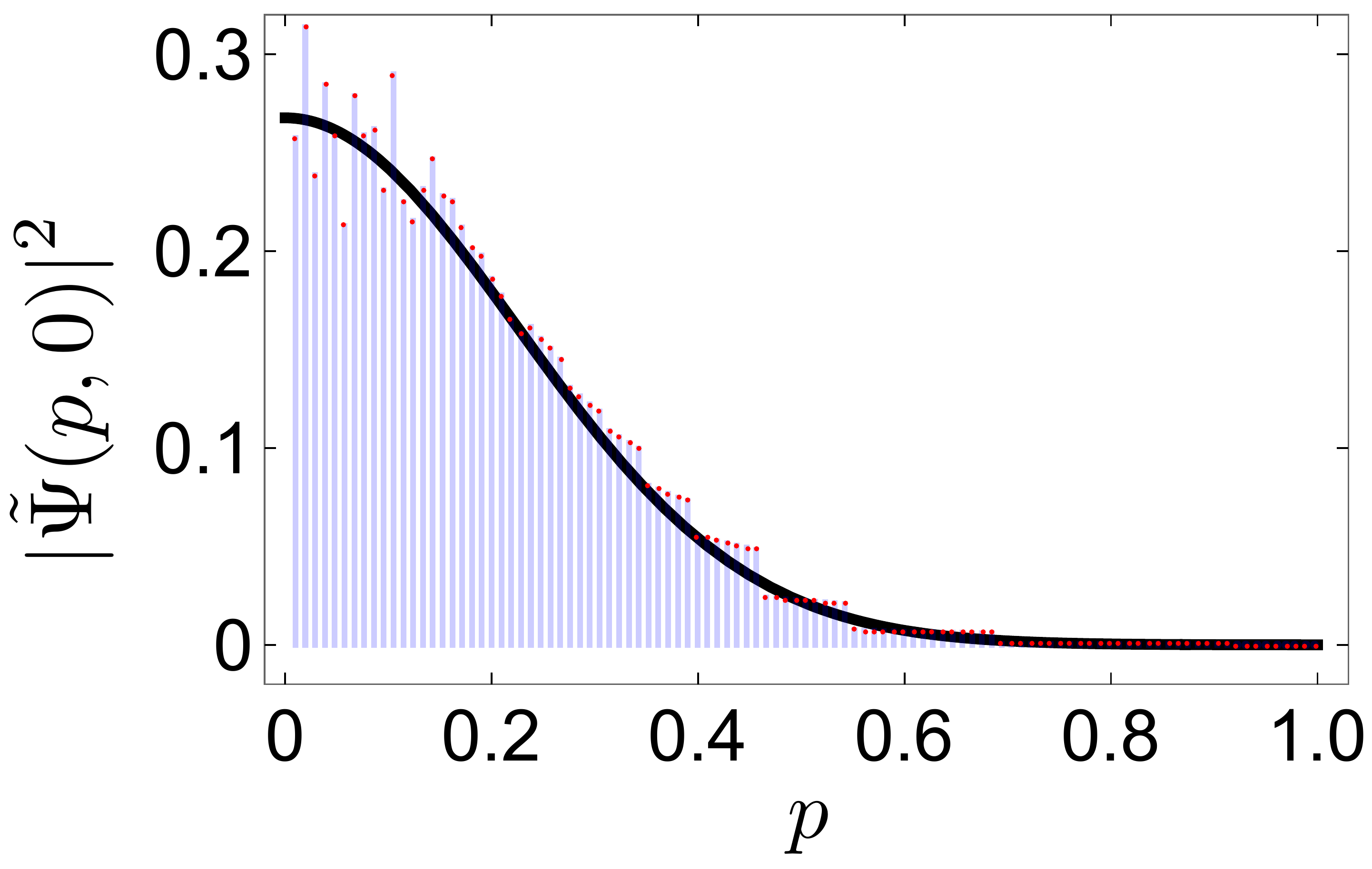}
\caption{\label{fig3b} 
The electron initial momentum distribution in a.u.\ extracted from fits of the histogram in Fig.\ \ref{fig3a} using the inverse IT Eq.\ (\ref{phiITsq}). The solid black curve shows the exact result $|\tilde \Psi(p,0)|^2$ from \eref{pwfn0} with the replacement $\sigma^2 \rightarrow \hbar/m \omega$ and $\omega = 0.1 \, \mbox{a.u.}$}
\end{figure}

Fig.\ \ref{fig3b} shows the initial $p > 0$ momentum distribution in atomic units extracted as a piecewise linear fit of the simulated arrival time detections using \eref{phiITsq}.
Evidently, the initial momentum distribution has been recovered. 
(Half the particles are emitted with $p < 0$ along the $-x$ axis and go undetected.)  

The increased errors in the extracted momentum distribution near $p=0$ are an artifact of finite sampling. Electrons released with small momenta require longer time of flights $t$ to reach a target $x$ compared with faster electrons. As is evident from Fig.\ \ref{fig3a}, there is a dearth of samples in the tail of the histogram for $t \gtrsim 10$ (besides being truncated at $t = 20$ out of numerical necessity) resulting in the small $p$ values being undersampled.

As noted in the previous section, an actual detector channel with a finite width of say $\delta x \approx 1 \, \mu\mbox{m}$ at $x= 30 \, \mbox{cm}$ will capture a bundle of trajectories with initial momenta in the range $\delta p \approx m \delta x/t$. With the average TOF $\bar{t} = 0.43 \, \mu\mbox{s}$ estimated above, one finds that $\delta x \, \delta \bar{p} \approx  m \, \delta x^2/\bar{t} \approx 0.02 \, \mbox{a.u.}$, which is well below the Heisenberg uncertainty limit with $\hbar =1$.

\appendix

\section{IT for unbound propagation in external fields}
\label{ITF}

In order to generalize the IT to wave packet propagation with macroscopic steering and extraction in external fields, we introduce the semiclassical propagator $K_{sc}(x, x';t)$ from Eq.\ (14) in the propagation integral Eq.\ (10). 
We assume for simplicity a time-independent hamiltonian so that only the total propagation time $t$ is relevant and the choice $t=0$ is arbitrary. 

 Again assuming only that $\Psi(x',0)$ is confined to a small range of $x'$ compared to asymptotic $\Psi(x, t \rightarrow \infty) $, one can expand the classical action $S_c(x, x';t)$ about $x'=0$ as
 \begin{equation}
\label{Sc0}
\begin{split}
S_c(x, x';t)& \approx S_c(x, 0;t) + \frac{\partial S_c}{\partial x'}\Big |_0\,x' \\
& = S_c(x, 0;t) - p'\,x',
\end{split}
\end{equation}
where $p' = -\partial S_c/\partial x'$ from Eq.\ (15). 
We then obtain for the asymptotic semiclassical propagator
 \begin{equation}
 \label{Ksc0}
K_{sc}(x, x';t) \approx K_{sc}(x, 0;t) \exp \left[-i p' x'/\hbar \right],
\end{equation}
form identical with Eq.\ (13).
With this approximate expression, the integral in Eq.\ (10) again evaluates directly as an inverse Fourier transform and gives 
\begin{equation}
\label{PsiITsc}
\Psi(x, t) \approx (2\pi \hbar)^{1/2} K_{sc}(x, 0;t) \, \tilde \Psi(p',0)
\end{equation}
but now generalized to arbitrary wave-packet extraction and steering in external fields.

As an example, we consider propagation and extraction with an applied external constant force $F$, for example a uniform electric field as deployed in a wide variety of spectrometry, or with a gravitational field in the case of gravity interferometry.
Again, the semiclassical and quantum propagators are identical with an action given by \cite{FBr2}
\begin{equation}
\label{S_F}
S_F(x, x';t) = F t x - \frac{F^2 t^3}{6m} + \frac{m}{2t} \left[x - x' -  \frac{F t^2}{2m} \right]^2,
\end{equation} 
which reduces to the free-motion action Eq.\ (11) in the $F=0$ limit.
We readily recover the classical trajectory for uniform accelerated motion, $x = x' + (p'/m)t + F t^2/2m$, with
\begin{equation}
\label{dSF_dx'}
-\frac{\partial S_F}{\partial x'}  = p' =  \frac{m}{t} \left[x - x' -  \frac{F t^2}{2m} \right] \approx \frac{m}{t} \left[x -  \frac{F t^2}{2m} \right],
\end{equation}
in the asymptotic $x \gg x'$ limit.
The Van Vleck density Eq.\ (16) remains, however, unchanged from free motion,
\begin{equation}
\label{VV1DF}
-\frac{\partial^2S_F}{\partial x\partial x'} = \frac{\partial p'}{\partial x} = \frac{m}{t} = \frac{dp'}{dx}.
\end{equation}
Then, the IT limit \eref{PsiITsc} becomes
\begin{equation}
\label{PsiIT_F}
\Psi_F(x, t) \approx \left(\frac{m}{i t}\right)^{1/2} e^{i S_F(x, 0; t)/\hbar}  \, \tilde \Psi(p',0),
\end{equation}
with $p'$ from \eref{dSF_dx'}.
Here, the action $S_F(x, 0; t)$ can be expressed compactly in terms of the final momentum defined by
\begin{equation}
\label{dSF_dx}
p = \frac{\partial S_F}{\partial x}  = p' + F t,
\end{equation}
with momentum transfer $F t$ from the field as required for uniform acceleration. Then,
$S_F(x, 0;t)  = -F^2 t^3/6m + p^2 t/2m$.

\section{IT from the stationary phase approximation} 
\label{ITSPA}

We derive the free-propagation IT for an \emph{arbitrary} initial momentum wave packet $\tilde \Psi(p,0)$ by evaluating the integral in Eq.\ (5) in stationary-phase approximation (SPA).  
The SPA in 1D is given by \cite{Schulman}
\begin{equation}
 \label{SPA}
\int_{-\infty}^{\infty}  g(p) \, e^{i \phi(p)/\hbar}  \, dp \approx g(p') \sqrt{\frac{2\pi i \hbar}{\phi''(p')}}  e^{i \phi(p')/\hbar}
\end{equation}
in the semiclassical limit $\phi \gg \hbar$, where $p'$ is a stationary phase point defined by $\phi'(p') \equiv 0$, and $g(p)$ a well-behaved function of $p$.

As $t \rightarrow \infty$, the exponent in Eq.\ (5) defines the rapidly oscillating phase $\phi(p) = p x - p^2 t/2m$ with stationary point $p \rightarrow p' = m x/t$ defined by $d\phi/dp = x - p t/m \equiv 0$. 
That is, asymptotically only $p$ values near the classical trajectories are relevant,
and $\phi(p') = m x^2/2t$ and $\phi''(p') = -t/m$. 
Therefore, in stationary-phase approximation, the integral in Eq.\ (5) evaluates approximately as 
\begin{equation}
\label{PsiFreeITSPA}
\begin{split}
\Psi(x, t) &\approx\left(\frac{m}{i t}\right)^{1/2} \exp\left[i \frac{m x^2}{2\hbar t}\right] \tilde \Psi(p',0)  \\ 
&= (2\pi \hbar)^{1/2} K(x, 0;t) \, \tilde \Psi(p',0),
 \end{split}
\end{equation}
in agreement with the IT limits Eqs.\ (8) and (12) but for arbitrary initial $\tilde \Psi(p',0)$ .


\begin{thebibliography}{99}
\bibitem{validity} The gaussian limit \eref{PsiFreeIT} in 1D is actually valid for all $x$. However, in the classical limit for $x \sim \sigma$ and asymptotic $t \gg T$, the momentum values $p = m x/t$  are negligible.
\end{thebibliography}
\end{document}